\newcommand{\hp}{\hat{\rm \bf n}}

\newcommand{\cmu}{\cos\theta}

\documentclass[useAMS, usenatbib, onecolumn]{mn2e}
\usepackage{graphicx}

\title[Gravitational wave astronomy of single sources with a pulsar timing array]
{Gravitational wave astronomy of single sources with a pulsar timing array}

\author[K.~J.~Lee et al.]{
K.~J.~Lee $^{1,2}$\thanks{Email: kjlee@mpifr-bonn.mpg.de}, 
N.~Wex$^{1}$, 
M.~Kramer$^{1,2}$,  
B.~W.~Stappers$^{2}$,  
C.~G.~Bassa$^{2}$, 
 \newauthor \
G.~H.~Janssen$^{2}$, 
R.~Karuppusamy$^{1,2}$
and
R.~Smits$^{2,3}$
\\
$^1${Max-Planck-Institut f\"ur Radioastronomie, Auf dem H\"ugel 69,
  D-53121 Bonn, Germany} \\
$^2${Jodrell Bank Centre for Astrophysics, University of Manchester,
  Manchester M13 9PL, UK} \\
$^3${Stichting ASTRON, Postbus 2, 7990 AA Dwingeloo, The Netherlands} \\
}

\begin{document}

\date{\today}
\pagerange{\pageref{firstpage}--\pageref{lastpage}} \pubyear{2010}
\maketitle
\label{firstpage}

\begin{abstract}

The stability of radio millisecond pulsars as celestial clocks allows
for the possibility to detect and study the properties of gravitational
waves (GWs) when the received pulses are timed jointly in a ``Pulsar
Timing Array'' (PTA) experiment. Here, we investigate the potential of
detecting the gravitational wave from individual binary black hole systems
using PTAs and calculate the accuracy for determining the GW properties.
This is done in a consistent analysis, which at the same time accounts
for the measurement of the pulsar distances via the timing parallax.

We find that, at low redshift, a PTA is able to detect the
nano-Hertz GW from super massive black hole binary systems with masses of
$\sim10^8 - 10^{10}\,M_{\sun}$ less than $\sim10^5$\,years before
the final merger.
Binaries with more than $\sim10^3 - 10^4$ years before merger are effectively
monochromatic GW, and those with less than $\sim10^3 - 10^4$\,years before
merger may allow us to detect the evolution of binaries.

For our findings, we derive an analytical expression to describe
the accuracy of a pulsar distance measurement via timing parallax.
We consider five years of bi-weekly observations at a precision of 15\,ns
for close-by ($\sim 0.5 - 1$\,kpc) pulsars. Timing twenty pulsars would
allow us to detect a GW source with an amplitude larger than $5\times
10^{-17}$. We calculate the corresponding GW and binary orbital parameters
and their measurement precision.  The accuracy of measuring the binary
orbital inclination angle, the sky position, and the GW frequency are
calculated as functions of the GW amplitude.  We note that the ``pulsar
term'', which is commonly regarded as noise, is essential for obtaining
an accurate measurement for the GW source location.

We also show that utilizing the information encoded in the GW signal passing
the Earth also increases the accuracy of pulsar distance measurements. If the
gravitational wave is strong enough, one can achieve sub-parsec distance
measurements for nearby pulsars with distance less than $\sim 0.5 - 1$\,kpc.

\end{abstract}


\section{Introduction}

Gravitational waves (GWs), ripples in space-time, perturb our four
dimensional space-time background, on which the pulsed radio radiation from
pulsars propagates. Through such space-time background perturbations, the GW
leaves its fingerprint in the arrival times of pulsar signals by introducing
an extra correlated component \citep{Sazhin78, Detweiler79, BCR84, Wahlquist87,
BH86}. By timing multiple pulsars (quasi-)simultaneously in a so-called
``Pulsar Timing Array'' (PTA) experiment, one can extract the GW signal,
even in the presence of various unrelated noise contributions in the timing
data \citep{RT83, HD83, BRN84, FB90, JLLW04, JHLM05, SV10, FL10}.

In general, GW sources are classified into two groups, the single
sources and a stochastic background component. A stochastic GW
background is a superposition of multiple single GW sources,
which are inseparable in frequency space \citep{CT02,
  Thorne89} due to finite frequency resolution of finite data.  The
techniques of detecting a single GW source and that of detecting a
stochastic background are therefore quite different. The detection
of a stochastic GW background using pulsar timing has already
been addressed in various studies \citep{Detweiler79, HD83,
  BH86, JHLM05, HJL09}. Here we concentrate on single GW sources.

Recently, many authors have drawn attention to the single source detection
problem \citep{JLLW04, SV10, BLF10, FL10, YHJ10, DF10, CC10}, because
of possible nearby { ($z\ll 1$)} coalescing binary systems that produce GW 
signals
detectable with PTAs \citep{WLH09, SVV09, Burke10}. In this paper,
we focus on the problem of detecting GWs from super-massive black
hole binaries (SMBHB), with masses from $10^8$\,$M_{\sun}$ to 
$10^{10}$\,$M_{\sun}$ and orbital frequencies in the nano-Hertz range, which 
makes
these SMBHBs indeed potential GW sources for PTA projects.  We will show
how PTAs can be used to measure the parameters of SMBHBs and their GWs.

The plan for the paper is as follows. We begin with a qualitative estimation of
the parameter space of the detectable SMBHB population for PTAs in
\S~\ref{sec:esti}. We then calculate the parameter estimation error for the GW
parameters in \S~\ref{sec:mear} where we describe the timing response to a
single GW source \S~\ref{sec:timres}, and present a careful and correct
treatment of the corresponding statistical problems \S~\ref{sec:zzbnd}.
Finally, we summarize and discuss our results in \S~\ref{sec:con}.

\section{The detectable black hole binary population}
\label{sec:esti}

In order to use the time-of-arrival (TOA) data from a PTA to detect GWs and
infer their parameters, two conditions have to be met.  First, the GW amplitude 
is
large enough {such that its statistic is significant to confirm the detection}, 
and secondly,
the GW frequency is in a frequency window where the pulsar timing technique
is sensitive. In this section, we will estimate the parameter space of the GW
sources, i.e.\ the parameter space of super massive black hole binaries
(SMBHBs), where they are observable to practical PTAs.

For a PTA, one needs at least five parameters to specify its
configuration, i.e.\ the number of pulsars ($N_{\rm psr}$) in the array,
the accuracy of the TOA ($\sigma_{\rm n}$), the total observing time
span ($T$), the duration ($\Delta T$) between two successive observing
sessions, and the pulsar distances $D_{\rm psr}$. These parameters
describe both the amplitude and frequency characteristics of a PTA.
{ For detecting a stochastic GW background, we do not need pulsar
distances \citep{JHLM05}. But as we will explain later, for a single
GW source, pulsar distances play an important role and need to be taken
into account in the analysis \citep{JLLW04, SVV09, FL10}.}

$T$ and $\Delta T$ determine the frequency characteristics of PTAs. For
example, the frequency resolution of a PTA is $\Delta f\approx 1/T$, and the
Nyquist frequency, the maximal recoverable frequency is given by $f_{\rm
max}\approx 1/(2\Delta T)$. In the standard pulsar timing data reduction
pipeline \citep{BH86, HJL09, LK05}, one fits for the periods
and period derivatives of pulsars. This removes the lowest-frequency components
in the pulsar timing data, and the minimal recoverable frequency $f_{\rm min}$
in the pulsar timing data is $f_{\rm min}\approx \Delta f =1/T$ \citep{JHLM05}.
$N_{\rm psr}$ and $\sigma_{\rm n}$ determine the amplitude
characteristics of the PTAs. One can detect a coherent sinusoidal signal among
all the pulsars in the PTA, if the amplitude of the signal is larger than
\citep{Scargle82} $\sigma_{\rm n}/\sqrt{N_{\rm psr} N_{\rm obs}}$, where
$N_{\rm obs}$ denotes the average number of independent TOA measurements for
each pulsar. The pulsar distance $D_{\rm psr}$ plays two roles in the whole
picture.  As we will show, it introduces another signal component in the pulsar
TOA data, i.e. the ``pulsar term'', which increases the signal-to-noise and also
gives a very long time baseline to investigate the GW evolution \citep{JLLW04}.

The mean GW amplitude $\bar{h}$ from a SMBHB averaged over the solid
angle of all orbital orientations of the binary, in units where $G = c
= 1$, is
\citep{Thorne89, WLH09}
\begin{equation}
	\bar{h} \equiv \sqrt{\langle |h_+|^2 + |h_\times|^2 \rangle} =      
	2\left(\frac{32}{125}\right)^{1/6} \frac{ {\cal M}^{5/3} \omega_{\rm g}^{2/3}
                (1+z)^{2/3}}{D_{\rm c}} \;,
  \label{eq:gwamp}
\end{equation}
where $h_+$ and $h_\times$ are the `plus' and `cross' gravitational-wave
amplitudes, and $\cal M$, $z$ and $D_{\rm c}$ are the chirp mass, the 
cosmological
redshift, and the co-moving distance of the binary. The $\omega_{\rm g}$ is the
GW angular frequency at the observer. In the standard cosmology model, the
co-moving distance $D_{\rm c}$ is

\begin{equation}
  D_{\rm c}=\frac{1}{H_{0}}\int_{0}^{z} 
  \frac{dz}{\sqrt{\Omega_{\Lambda}+\Omega_{\rm M} (1+z)^3}} \;.
  \label{eq:covdis}
\end{equation}
where we used the dark energy density $\Omega_{\rm \Lambda} = 0.7$, matter 
density $\Omega_{\rm M} = 0.3$, and the Hubble constant $H_{0}=73$\,${\rm km 
s^{-1}\,Mpc^{-1}} = 2.37 \times 10^{-18}$\,${\rm s^{-1}}$ \citep{PDG08}. The 
angular frequency relates to the black hole binary chirp mass and the time 
before final merger of the binary via \citep{Hughes09}
\begin{equation}
	\omega_{\rm g} = 2\left(\frac{5}{256}\right)^{3/8}
	                 \frac{1}{ {\cal M}^{5/8} t_{\rm m}^{3/8} (1+z)^{5/8}} \;,
  \label{eq:freq}
\end{equation}
where $t_{\rm m}$ is the time before final coalescence of the binary system
at the observer, i.e.\ the time span between the time of `present' to the time of the
binary's final merger. The GW radiation takes away the energy of the 
SMBHB, which induces a frequency chirping GW. The angular frequency of the GWs
increases at the rate of $\dot{\omega}_{\rm g}$, where 
\begin{equation}
	\dot{\omega}_{\rm g} = \frac{3}{4}\left(\frac{5}{256}\right)^{3/8}
	      \frac{1}{{\cal M}^{5/8} t_{\rm m}^{11/8}(1+z)^{5/8}} \;.
  \label{eq:fdot}
\end{equation}

The GWs from a black hole binary introduce signals in the pulsar TOA data. It
turns out that the GW induced timing signal $R(t)$ depends on both the GW
strain at the pulsar and the GW strain at the Earth, i.e. $R(t) \sim
\left[h(t,0) - h(t-D_{\rm psr}, {\rm \bf D_{\rm psr}})\right]/2\omega_{g}$
\citep{EW75} \footnote{Here, we ignore the unimportant geometrical factor for
this qualitative estimation.}. The $h(t,0)$, the \emph{Earth term}, is the GW
strain at the Earth at the time when we receive the pulse; while $h(t-D_{\rm
psr}, {\rm \bf D_{\rm psr}})$, the \emph{pulsar term}, is the GW strain at the
pulsar at the time when the pulse was emitted. Thus, in principle, the GW from
a quasi-circular SMBHB introduces two quasi-monochromatic components in the
pulsar timing residuals. One of the components comes from the Earth term, the
other comes from the pulsar term. Due to the evolution of SMBHB system, the
frequency of the pulsar term and the Earth term are different. The frequency
difference between the two components is $\delta f \approx \dot{\omega}_{\rm g}
D_{\rm psr} /2\pi$. 

{To be observable, the GW induced timing signal has to be large enough compared
to the other noises affecting the TOA accuracy. For most of the GW single 
sources, they evolve slowly ($\delta f \ll \omega_{\rm g}/2\pi$), thus we can 
ignore the frequency difference between the Earth and the pulsar term and use a 
single frequency to calculate the amplitude of the induced pulsar timing signal, 
i.e.}  \begin{equation}
	\frac{\bar{h}}{2\omega_{\rm g}} \ge \frac{\sigma_{\rm n}}{\sqrt{N_{\rm psr} 
	N_{\rm obs}}} \;,
\label{eq:estisig}
\end{equation}
{which is generalized to red noise in Appendix~\ref{sec:genred}.}  Consequently,
\begin{equation}
  {\cal M}^{15/8} t_{\rm m}^{1/8} \ge 
      \frac{ 5^{5/8} }{ 2^{3/2} } \,
      \frac{ \sigma_{\rm n} D_{\rm c} }
           { (N_{\rm obs} N_{\rm psr})^{1/2} (1+z)^{7/8} } \;.
 \label{eq:amplim}
\end{equation}

We also need to check if the GWs meet the frequency range of PTAs, i.e.\ 
$f_{\rm min}\le\omega_{\rm g}/2 \pi\le f_{\rm max}$, which leads to
\begin{equation}
  \frac{ 5^{3/8} }{ 8\pi } \, \frac{ T }{ (1+z)^{5/8} } 
  \ge {\cal M}^{5/8} t_{\rm m}^{3/8} \ge 
  \frac{ 5^{3/8} }{ 4\pi } \, \frac{ \Delta T }{ (1+z)^{5/8} } \;.
  \label{eq:freqlim}
\end{equation}
equation.~(\ref{eq:amplim}), the amplitude condition, and the
equation.~(\ref{eq:freqlim}), the frequency condition, together determine the 
SMBHB
parameter space, in which the GWs from such SMBHB will be observable to the
PTA.

It is interesting to see when evolutionary effects become important as 
discussed by \cite{JLLW04}. If the frequency difference between the pulsar and 
the Earth term is larger than the frequency resolution of the PTA, one can 
detect the evolution of the SMBHB. This requires $\delta f\ge \Delta f$
\citep{Seto02}, i.e.
\begin{equation}
  {\cal M}^{5/8} t_{\rm m}^{11/8} \le
      \frac{3}{8\pi}\left(\frac{5}{256}\right)^{3/8}
      \frac{T D_{\rm psr}}{ (1 + z)^{5/8}} \;.
  \label{eq:evolim}
\end{equation}

It is also interesting to examine when the GWs from SMBHBs are no longer
quasi-monochromatic. This is the case, when the SMBHB arrives at the dynamical
evolution phase, as the orbital radius approaches the innermost stable
circular orbit (ISCO). The quasi-monochromatic description of the GW is
therefore only valid before the binary arrives at the dynamical phase, i.e.\ the
ISCO is a definitive upper bound for the frequency range of the
quasi-monochromatic regime. For an equal mass binary the orbital angular 
frequency of
the ISCO is given by $\omega_{\rm orb}=\omega_{\rm g}/2 \simeq 
1/(6^{3/2}2^{6/5}{\cal M}) $
\citep{BI03}. Consequently
\begin{equation}
  \frac{ t_{\rm m} }{ {\cal M} } \ge \frac{405}{2^{4/5}} \, (1 + z) \;.
  \label{eq:lastlim}
\end{equation}

\begin{figure}
  \centering \includegraphics[totalheight=2.4in]{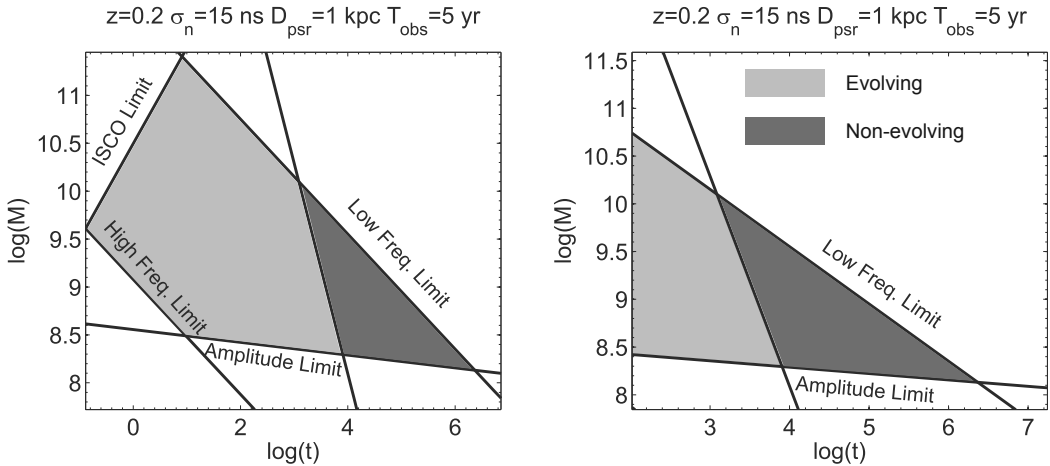}
        \caption{The parameter space of SMBHBs as detectable GW
          sources for a PTA. The right panel shows an enlarged region
          of the left panel. The x-axis is the time span from the
          `present' to the final merger in years; the y-axis is the
					chirp mass of the SMBHB system in Solar masses.  The corresponding 
					limits are labeled on the line, i.e. `Amplitude Limit' for 
					equation.~(\ref{eq:amplim}), `Low Freq. Limit' and `High Freq. Limt' 
					for
					equation. ~(\ref{eq:freqlim}), and `ISCO Limit' for 
					equation.~(\ref{eq:lastlim}).  The parameter space for evolving and 
					non-evolving GW sources is shaded using using light gray and dark 
					gray respectively as indicated in the figure.  It should be noted
					that, although the dark gray region may look smaller than the
					light gray region, the \emph{dominant detectable SMBHB
						population belongs to the dark gray region}, where we
          encounter smaller mass and longer rest lifetimes. For our
          calculations we considered SMBHBs with a and assumed
          bi-weekly observations of a PTA with 20 pulsars, of which
          the parameters are shown above the
          panels. } \label{fig:esti}
\end{figure}

We summarize the results in Figure~\ref{fig:esti}. We can see that PTAs will be
sensitive to the SMBHB population with a mass range from $10^{8}$ to $10^{10}\,
M_{\sun}$ and about $10^{5}$ years before the merger. For those SMBHBs with
$t_{\rm m}$ less than $\sim 10^3 - 10^4$\,years, the frequency chirp will be
visible in the 5-year data. In Figure~\ref{fig:esti}, the non-evolving SMBHBs
are believed to be the dominant population in terms of numbers, since they have
lower masses (10 times smaller) and longer lifetimes ($10^2 - 10^3$ times
longer). Hence, we will focus on the non-evolving sources in the rest of the
paper. The discussion for the evolving GW sources, i.e. the source in the
starred region of Figure~\ref{fig:esti}, will be presented elsewhere. The
SMBHBs approach the last stable orbit only about a few years before merger, so
that for all practical PTA observations, we do not need to consider the
dynamical merging phase -- this is in contrast to the case of ground based GW
detectors operating at much higher frequencies.

\section{Ability of a PTA to measure the parameters of a non-evolving SMBHB}
\label{sec:mear}

In the previous section, we estimated the parameter space for the
PTA-detectable SMBHB population. In this section, we turn to a detailed 
analysis of the GW parameter estimation problem. We firstly calculate the
pulsar timing response to a monochromatic GW, before we determine the
statistical error of estimating GW parameters.

\subsection{Pulsar timing signal with a monochromatic GW}
\label{sec:timres}

Signals in pulsar TOA data come from many different origins, which
include the motion of the Earth, the motion of the pulsar (in particular if it
is in a binary system), changing relativistic delays in the signal propagation,
the scattering, and the dispersing due to space plasma, the gravitational
waves in the space time background, etc.\ \citep{MT77, BH86, Stairs03, LK05}.
We are, here, particularly interested in the timing signal components related
to gravitational waves.

A persistent GW introduces two different signal components in TOA
data, i.e.~the ``Earth term'' and the ``pulsar term'', where the Earth
term depends on the GW strain at the Earth and the pulsar term
contains the GW strain at the pulsar.  Due to the phase coherence, the
Earth term and the pulsar term interfere with each other depending on
the pulsar distance, the pulsar direction, and the gravitational wave
propagating direction. Because of such pulsar-Earth term interference,
the pulsar distance also becomes an important variable in the GW
detection problem. The most relevant pulsar timing signals are then
the GW induced signal $R_{\rm g}$, the pulsar timing parallax $R_{\rm
  par}$, and the noise $n$ which includes other noise contributions to
the TOAs.  Both GW-induced signal and timing parallax depend on
  the pulsar distances. The timing residual signal can be written as
\begin{eqnarray}
	R(t) = T_{0}+P_0 N +\frac{1}{2} P_{0} \dot{P_0} N^2+ \Delta_{\sun}+R_{\rm 
	g}(t) + R_{\rm par}(t) + n(t|\sigma_{\rm n}) \;,
  \label{eq:fullsig}
\end{eqnarray}
{ where the $N$ is the sequential pulse number, the $T_{0}, P_0, \dot{P_0}$ is 
the initial epoch, period and period derivative for the pulsar, while 
$\Delta_{\sun}$ represents the standard Solar system correlations 
\citep{LK05}.  }The $n(t|\sigma_{\rm n})$ is taken to be white zero-mean
Gaussian random variables with an RMS level of $\sigma_{\rm n}$
to simulate the other noise contributions to the TOA. {In our calculation, we 
fit for $P_0$, $\dot{P_0}$, and pulsar sky position using a least-square 
fitting as in standard pulsar timing
pipelines \footnote{If the pulsar is in a binary system,
the fitting of the orbital motion will also remove some signal power.  } 
\citep{LK05}}.

{ As shown in Appendix~\ref{sec:2dres}, the pulsar timing signal induced by a 
single GW source is} \begin{eqnarray}
  R_{\rm g}(t) &=& \frac{h_0}{2\omega_{\rm g}}\,
                   \frac{\sin(\Delta\Phi/2)}{1 - \cos\theta} 
                   \left\{ \left[B_1\cos(2\phi) + B_2 \sin(2\phi)\right] 
                   \cos(\omega_{\rm g}t - \Delta\Phi/2) (1 + \cos^2\iota) \right.
  \nonumber \\
  && \left.+2\left[B_2 \cos(2\phi)-B_1 \sin(2\phi)\right]
     \sin(\omega_{\rm g}t-\Delta\Phi/2)\cos\iota \right\} \;,
  \label{eq:gwsig}
\end{eqnarray}
where the $B_1$, $B_2$ and $\Delta\Phi$ depend on the geometrical 
configuration of the pulsar and GW position by
\begin{eqnarray}
  B_1 &=& (1 + \sin^2\beta) \, \cos^2\beta_{\rm p} \cos[2(\lambda -\lambda_{\rm p})]
			 - \sin(2\beta) \sin(2\beta_{\rm p}) \cos(\lambda - \lambda_{\rm p})
			 + (2 - 3 \cos^2\beta_{\rm p}) \cos^2\beta  \;, \\
  B_2 &=& 2 \cos \beta  \sin (2 \beta_{\rm p} ) \sin(\lambda - \lambda_{\rm p}) -
          2 \sin\beta \cos^2\beta_{\rm p} \sin [2 (\lambda - \lambda_{\rm p})] , \\
	\Delta\Phi &=& \omega_{\rm g} D_{\rm psr} (1 - \cos\theta) \;.
	\label{eq:phasd}
\end{eqnarray}
{Here, the $\lambda$ and $\beta$ are the ecliptic longitude and latitude of the 
GW source, the $\lambda_{\rm p}, \beta_{\rm p}$ are the ecliptic longitude and 
latitude for the pulsar position, the angle $\iota$ is the orbital inclination 
of the GW binary source, i.e.\ the angle between orbital angular momentum and 
the direction to the Earth. The $h_0 = 2^{4/3} {\cal M}^{5/3}\omega_{\rm
g}^{2/3} D_c^{-1} (1+z)^{2/3}$ is the GW amplitude. }

For a persistent GW source, the pulsar term and the Earth term are
coherent and have a phase difference of $\Delta\Phi$. Such
pulsar-Earth interference leads to two consequences: 1) it changes the
amplitude of the timing signal; 2) it changes the phase of
the timing signal. One can see the two effects from
equation.~(\ref{eq:gwsig}). The $\sin(\Delta\Phi/2)$ term is the modulation
for the signal amplitude, while the $\Delta \Phi/2$ in the term
$\omega_{\rm g} t - \Delta\Phi/2$ is the signal phase shift (see
Appendix \ref{sec:2dres} for the details).

\begin{figure}
\centering
\includegraphics[totalheight=3.0in]{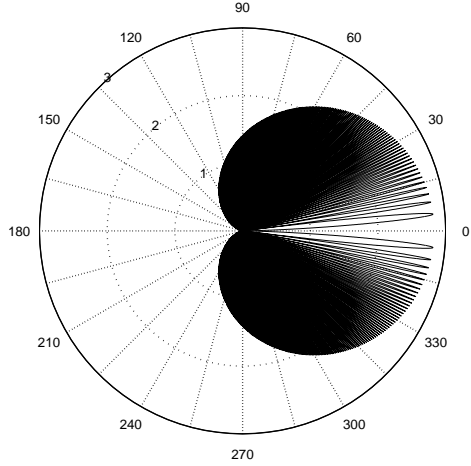}
\caption{Response pattern $H$ of a single-pulsar timing response to a
  single monochromatic GW source. For illustration purposes, the
  pulsar distance is chosen to be small with a value of 200\,pc and
  the GW period is chosen as 5 years, in order to show the structure
  of the response pattern. The GW source is in the $0^\circ$ angle
  position, the orbital plane inclination is $90^\circ$ and the orbital
  plane coincides with the paper plane. In this way the plotted response pattern 
	is, in fact, the term $(1+\cos\theta)\sin(\Delta\Phi/2)$ in 
	equation.~(\ref{eq:2dtim}).}
\label{fig:beampat}
\end{figure}

The pulsar timing response $H$ to the GW signal is the normalized ratio between
the amplitude of GW-induced pulsar timing signal and the
amplitude of the GW, i.e.
\begin{eqnarray}
  H &=& \frac{ (\textrm{GW induced amplitude in TOA}) \times 
               (\textrm{GW angular frequency}) }{ (\textrm{GW amplitude}) }
        \nonumber \\ 
    &=& \frac{\omega_{\rm g}}{h_0}
        \left[\lim_{T\to \infty}\left(
        \frac{1}{T}\int_{0}^{T} R_{\rm g}^2(t) \, dt \right)\right]^{1/2} 
    \;=\quad \frac{\sin(\Delta\Phi/2)}{4(1 - \cos\theta)}\sqrt{B_3}, 
  \label{eq:resH}
\end{eqnarray}
where
\begin{eqnarray}
  B_3 &=& (B_1^2 + B_2^2)(1 + 6\cos^2\iota + \cos^4\iota)
		+\left[ (B_1^2 - B_2^2)\cos(4\phi)+2B_1 B_2\sin(4\phi) \right] 
		\sin^4\iota.
\end{eqnarray}
The response $H$ is a function of the GW source position, the pulsar location,
and the orientation of the SMBHB orbit. Due to the interference between the
pulsar term and the Earth term, the response $H$ also depends on the distance
of the pulsar through the term $\sin(\Delta\Phi/2)$ in 
equation.~(\ref{eq:resH}).
Figure~\ref{fig:beampat} shows the timing response as a function of $\theta$, 
the
angle between pulsar direction and GW source direction. The response
$H(\theta)$ is simply the response pattern for a single-pulsar
GW detector. As shown in Figure~\ref{fig:beampat}, the response pattern is made
up by a large number of spiky lobes. The envelope of these lobes depends on the
GW source position and polarization, and the spiky lobes are due to the
interference between the pulsar and the Earth term. Given the angular frequency
of the GW $\omega_{\rm g}$ and the pulsar-Earth distance $D_{\rm psr}$, one can
show that the angular width of the each lobe is approximately $\theta_{\rm
lobe} \sim 1/(2 D_{\rm psr} \omega_{\rm g})$ ignoring the weak dependence on
$\theta$. For practical PTA purposes, the $D_{\rm psr}\sim 1$\,kpc and
$\omega_{\rm g}\sim$\,yr$^{-1}$, so $\theta_{\rm lobe}\sim 10^{-3}$ rad.  It is
the existence of these spiky lobes which gives PTA experiments the power to
accurately locate single GW sources, given that the pulsar distances are well
measured.

By measuring the phase difference $\Delta \Phi$ between the pulsar
term and the Earth term and by comparing data from different pulsars,
the pulsar distance can be inferred. However, the result of such a
calculation would have multiple solutions, i.e.\ we cannot
discriminate between the case of $\Delta \Phi$ and the cases of any
$\Delta \Phi+4 n \pi$, where $n=0,1,2,\dots$, since we have shown in
equation.~(\ref{eq:gwsig}) that the GW-induced timing signal $R_{\rm g}$ is
a periodical function of $\Delta \Phi$ with a periodicity of $4\pi$.
In this way, one can only measure the modulo of the pulsar distance by
$4 \pi\omega_{\rm g}^{-1} (1-\cos\theta )^{-1}$.  On the other hand,
if there is another, independent way of measuring the pulsar distance, 
even with a lower but sufficient accuracy to remove such confusion
about multiple solutions, one can use the GW signal to increase the
accuracy of pulsar distance measurements by utilizing the GW distance
modulo.

For millisecond pulsars, such an independent distance measurements is
available from the timing parallax \citep{BH86, RT91, LK05, VLM10}. It
is expected that with future radio telescopes, such as the SKA
\citep{KS10, STWKS10}, timing parallax measurements will yield very
precise distances for nearby pulsars.  Indeed, SKA timing parallax measurements
promise to be of sufficient accuracy to help us to identify the
correct solution for $\Delta \Phi$  and, hence, to measure the pulsar
distance with even better precision.

In order to consider this further, we study a simplified version of the
timing parallax term. {In reality, the timing parallax is derived from  the 
detailed motion of the Earth from Solar System
dynamics \citep{Almanac10}.  For the purpose of this paper, it is sufficient to 
keep the leading term of the timing parallax, i.e.\ assuming a circular motion 
of the Earth,}
\begin{equation}
	R_{\rm par}(t) = \frac{ \cos[2(\lambda_{\rm psr}-\lambda_{\earth}(t)) ]
                          \cos^2\beta_{\rm psr} r_{\earth}^2}{4D_{\rm
                          psr}} \;,
  \label{eq:parsig}
\end{equation} 
where the term $r_{\earth}$ is the average distance between the Sun
and the Earth, and $\lambda_{\earth}(t) = 2\pi (t/1\,{\rm year})$ is the
ecliptic longitude of the Earth at time $t$.  This form of
timing parallax assumes that the eccentricity of the Earth orbit is
zero. This assumption is valid for cases where the pulsar is not too
close to the ecliptic poles, i.e.~($-89^\circ \le \beta \le
+89^{\circ}$), {such that a timing parallax signal is not dominated by the 
Earth orbit's eccentricity}.  As this will generally be the case, the error of 
the measured pulsar timing parallax distance is (see
Appendix~\ref{sec:estipar} for details)
\begin{equation}
  \sigma_{ D_{\rm psr} } =
    \frac{4\sqrt{2} \, \sigma_{\rm n}D_{\rm psr}^2}{\sqrt{N_{\rm obs}}
    \, r_{\earth}^2 \cos^2\beta_{\rm psr}}
	\simeq  \frac{2.34}{\cos^2\beta_{\rm psr}} \left(\frac{N_{\rm 
	obs}}{100}\right)^{-\frac{1}{2}}
    \left( \frac{D_{\rm psr}    }{ 1\,{\rm kpc} } \right)^2 \left(
		\frac{\sigma_{\rm n} }{ 10\,{\rm ns}  } \right) \, {\rm pc} \;,
  \label{eq:distmeasure} \end{equation}
where $N_{\rm obs}$ is the number of TOAs. The numerical factor is derived 
assuming that the time span of pulsar data is longer than one year. {In a
real data analysis, one always uses the full Solar System ephemeris.} We 
compared
equation.~(\ref{eq:distmeasure}) with results from numerical simulations
based on TEMPO\footnote{See
  http://www.atnf.csiro.au/research/pulsar/tempo/.} and the planetary
ephemeris DE405 \citep{Standish98}.  For pulsars with $-89^{\circ} \le
\beta \le +89^{\circ}$, we find that the simplified version of the
timing parallax shown above agrees with the correct result derived
from TEMPO within a few percent difference, justifying the usage of
equation.~(\ref{eq:distmeasure}) for the purpose of the present paper. We
note that the validity of equation.~(\ref{eq:parsig}) comes from the fact
that the Earth orbital eccentricity is small and that we are
investigating measurement accuracies, where the effect of orbital
eccentricity is of even higher order. According to
equation.~(\ref{eq:distmeasure}), with a timing accuracy at the 10 to 30-ns
level, one can use the timing parallax to measure the pulsar distance
accurately to a few light years for pulsar distances of less than 1\,kpc. This 
distance accuracy become comparable to the wavelength of the
GW, and the timing parallax measurement is therefore indeed a
potential technique to remove the pulsar distance
confusion. Both GW parameters and pulsar distances should thus
be estimated from pulsar timing data at the same time. In the
following, we estimate the corresponding accuracy of the GW parameters
and pulsar distances measurements based on the signal timing of
equation.~(\ref{eq:fullsig}).

\subsection{Vector Ziv-Zakai bound for signals with additive white Gaussian 
noise}
\label{sec:zzbnd}

{We are, now, going to determine the statistical error of estimating GW 
parameters using data from a PTA.} A well known and popular statistical 
technique to calculate such lower
bounds of the statistical accuracies of parameter estimators is the
Cramer-Rao bound (CRB) based on the Fisher information
\citep{Fisz63}. However CRB is also known for predicting a too small
value, which can not be achieved in practical cases, especially, when
the signal-to-noise ratio (SNR) is low \citep{CZZ75, BSE94,BEv96,NV97,
  TB97}.  Besides the low-SNR problem, the CRB is derived in the local
sense, which assumes that an unbiased estimator (UBE) exists
\citep{Fisz63}. Due to this assumption, the CRB is not applicable to
cases, where the UBE does not exist \citep{Jaynes03}. For example,
CRB predicts smaller-than-achievable bounds, when there are multiple
isolated regions with similar likelihoods in the parameter
space. In the present GW parameter estimation problem, the
pulsar distance confusion introduces multiple equal-likelihood
structures for both pulsar distances and GW source locations at low
SNR. This urges us to use, instead of CRB, other techniques to
overcome these difficulties.
Many statistical bounds were investigated in the passed
decades (see \cite{TB97} for a review). It turns out that a group of
statistical bounds, which originates from \cite{CZZ75}, gives more
reliable bounds than the CRB; and these bounds approach the
CRB when the SNR is high and multiple equal-likelihood
confusions are resolved. These bounds are now referred to as Ziv-Zakai
(ZZ) bounds.  In this paper, we adopt the ZZ bounds to investigate the
error performance of the GW parameter estimators. We also calculate
the same results using CRB for comparing purposes.

For a $k$-dimensional vector parameter $\lambda_{i}$ ($i=1\ldots k$),
and its estimator $\hat{\lambda}_{i}$, the estimation error is
quantified by the correlation matrix ${\cal C}_{ij} = \langle
(\hat{\lambda}_{i} - \lambda_{i})(\hat{\lambda}_{j} - \lambda_{j})
\rangle$. The error for the estimator $\hat{\lambda}_{i}$ is just the
diagonal part of the correlation matrix, i.e. $\sigma_{\lambda_i}^2 =
{\cal C}_{ii}$. The ZZ bound states that the following inequality
holds
\begin{equation}
  \kappa_i{\cal C}_{ij}\kappa_j \ge 
    \frac{1}{2} \int_{\Omega_{\eta}} {\cal V} \left\{
    \max_{\kappa_{i}\delta\lambda_{i}=\eta} \left[ 
    \int [P_{\lambda_{0,i}}(\lambda_i) +
          P_{\lambda_{0,i}}(\lambda_i + \delta\lambda_i)]
          P_{\epsilon,{\rm min}}(\lambda_i, \lambda_i+\delta\lambda_{i}) \,
          d\lambda_i \right]\right\} \eta \, d\eta \;,
  \label{eq:zzb}
\end{equation}
where $\kappa_i$ can be any constant
vector. Following the Einstein summation convention, we sum over the 
index if it appears twice in a single term. The 
$P_{\lambda_{0,i}}(\lambda_{i})$, $P_{\lambda_{0,i}}(\lambda_i+\delta\lambda_i)$ 
are the prior probabilities for parameters $\lambda_{i}$ and 
$\lambda_{i}+\delta\lambda_{i}$ given that the true value of the parameters are 
$\lambda_{0,i}$. The $\delta\lambda_{i}$ is a $k$-dimensional vector, which 
maximizes the integral in the rectangular bracket under the constrain that 
$\kappa_{i} \delta\lambda_{i}=\eta$ in equation.~(\ref{eq:zzb}), where $\eta$ is 
the scalar parameterizing the integral path.  The $P_{\epsilon,{\rm 
min}}(\lambda_i, \lambda_i + \delta\lambda_i)$ is the minimal probability of 
making an error, when discriminating between the parameter set $\lambda_i$ and 
$\lambda_i+\delta\lambda_i$. The scalar function $\cal V$ is the `valley-filling 
function' defined as
\begin{equation}
  {\cal V}[f(\eta)]=\max_{\eta'\ge 0}f(\eta+\eta') \;.
  \label{eq:valley}
\end{equation}
The integration domain $\Omega_\eta$ is determined by prior
information. For example, we may use the fact that the PTA pulsars are
all located in the Galaxy, which allows us to confine the integrations
for the pulsar distances to the radial size of the Galaxy.  We refer
to \cite{BSE94} for details of the formalism in the above. If
  $\kappa_i$ is taken to be the $k$-dimensional unit vector,
	equation.~(\ref{eq:zzb}) gives the bounds for the parameter estimation
  errors. We present more details on how we calculate the Ziv-Zakai
  bound for our problem in Appendix~\ref{sec:zzbndapp}.

\begin{figure}
	\centering \includegraphics[totalheight=6.0in]{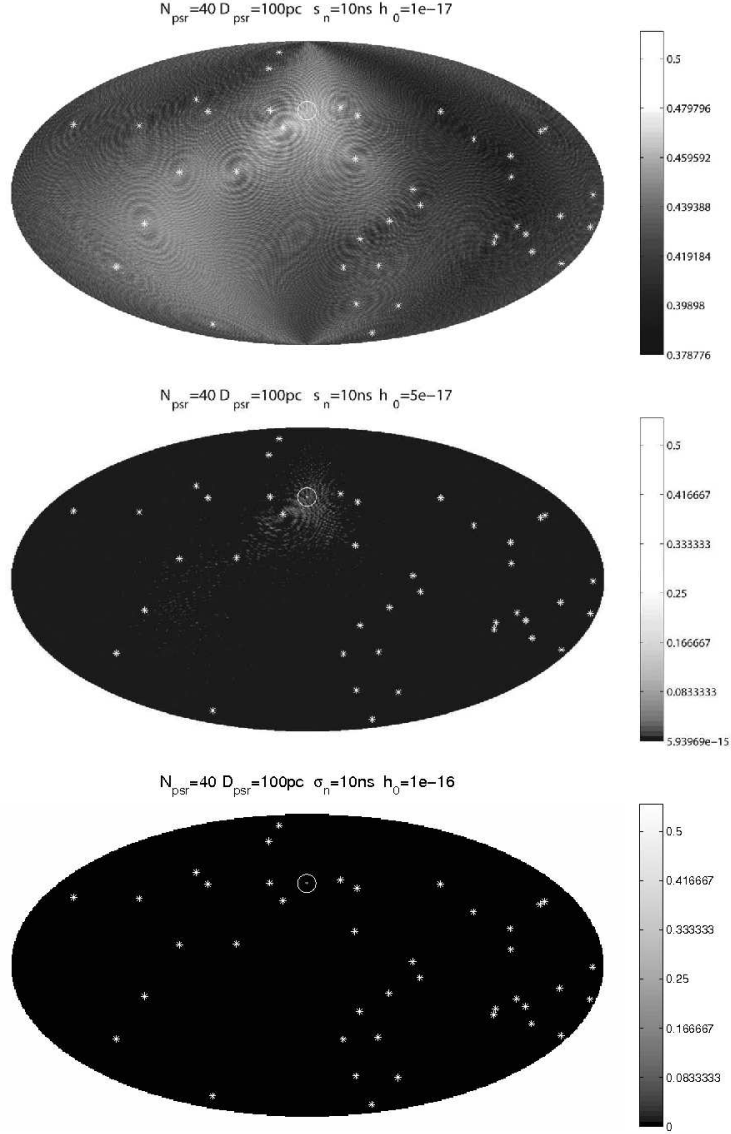}
	\caption{The minimal
	probability of incorrectly determining the source location. Here,
  the GW source is located at the center of the white circle.
	The gray scale shows the minimal probability of making a mistake
	($P_{\epsilon, {\rm min}}$) when deciding between the true GW
	source location and any other location in the $\lambda-\beta$
	parameter space.  When the GW amplitude is low (top panel),
	$P_{\rm \epsilon, min}$ is large ($\simeq 0.5$) over the whole
	sky. When the GW amplitude is strong enough (bottom panel), the
	probability of determining an incorrect GW location is lower and
	the multiple-solution confusion is removed. Thus only the real GW position is 
	left. For intermediate GW
	amplitudes (middle panel), there are multiple isolated positions,
	which all have high probability of being indistinguishable from
	the true GW source position. For the illustration purpose here,
	these figures are calculated for a PTA with 40 pulsars, where
  each pulsar has a 10-ns timing accuracy and a distance of 100\,pc. The GW 
  frequency is taken to be $1.2$ year.  The white
	stars indicate the locations of the PTA pulsars. The visible
  interference pattern is due to the response pattern discussed in
	the text. } \label{fig:skymap}
  \end{figure}

In order to demonstrate why the ZZ bound is more trustful than
classical CRB, we take the GW source position as an example. In
Figure~\ref{fig:skymap}, we show the $P_{\epsilon, {\rm min}}$ as a
function of GW source location. For the case when the GW amplitude is
low, there is a high probability of making a mistake in discriminating
between the true GW source location and other locations in multiple
isolated regions. The CRB is flawed by the large local derivative near
the most probable value and it thus underestimates the error by
ignoring other isolated high likelihood regions. The ZZ bound, unlike
CRB, is an average of the parameter distance $\eta$ weighted by the
error probability $P_{\rm \epsilon, min}$ over the entire parameter
space. Thus, we only consider the ZZ bound to be suitable for a
correct treatment of cases with potential multiple solution confusion.

\begin{figure}
	\centering
	\includegraphics[totalheight=3.0in]{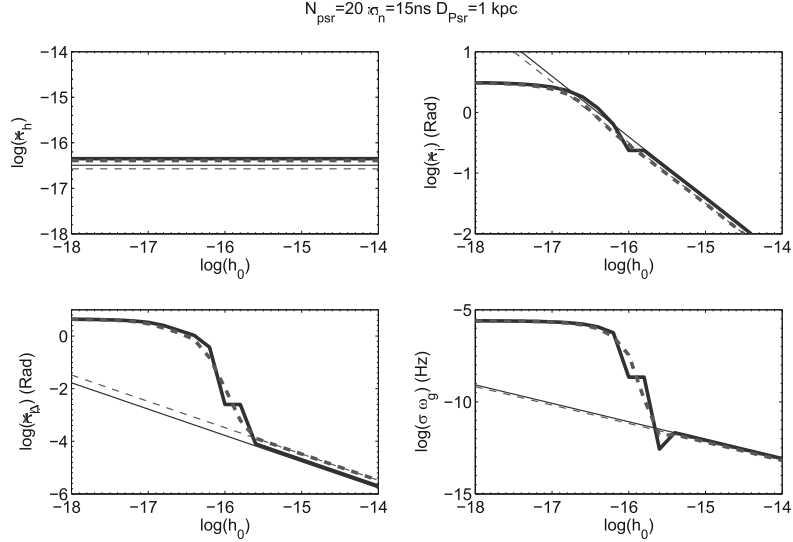}
	\caption{The parameter estimation error as a function of GW
          amplitude.  $\sigma_{\rm h}$ is the lower bound for the
          accuracy in estimating the GW amplitude $h_0$, $\sigma_{\rm
            i}$ is for the orbital inclination angle of the GW source,
          $\sigma_{\rm \Delta}=\sqrt{ \sigma_{\lambda}^2\sin^2\beta +
            \sigma_{\beta}^2}$ is the lower bound for the accuracy in
		  measuring the GW source position, where $\sigma_{\lambda}$
          and $\sigma_{\beta}$ are the errors for estimating $\lambda$
          and $\beta$, respectively. The $\sigma_{\rm \omega_{\rm g}}$
          is the lower bound of the measurement accuracy of the GW
          frequency. The $h_{0}$ is the GW amplitude. One can see that
          the error of amplitude estimation is not sensitive to the
          true GW amplitude, while the uncertainties in the other
					parameters decrease as the GW amplitude becomes larger \citep{TS02}.  
					The
          thick lines are the results from the ZZ bound, and the thin
          lines are the bounds calculated from the CRB.
          The dashed line is for a GW with a period of 2.2\,years, while the 
          solid
          line is for a GW with a 4.4\,year period. The total observation span 
          $T$ is 5 years and the duration $\Delta T$ between two successive 
          observation are 2 weeks. The pulsars in the PTA
          are uniformly distributed over the sky at equal distance.
          Details about the PTA, such as the number of pulsars, timing
          precision and pulsar distances are given at the top of the figure.}
	\label{fig:hdep}
\end{figure}

  Since we are interested in calculating the errors of the estimated
  GW source parameters as functions of the GW and pulsar parameters,
  the priors $P_{\lambda_{0,i}}(\lambda_{i})$ and
  $P_{\lambda_{0,i}}(\lambda_{i} + \delta\lambda_{i})$ take forms of
  Dirac $\delta$-functions, i.e.  $P_{\lambda_{0,i}}(\lambda_{i})
  =P_{\lambda_{0,i}}(\lambda_i + \delta\lambda_i)=
  \delta(\lambda_i-\lambda_{0, i})$, where $\lambda_{0,i}$ is the true
  parameter. With the pulsar signal model specified by
	equation.~(\ref{eq:fullsig}), one can calculate the ZZ bound for the
  parameter estimators, i.e. for the pulsar distance $D_{\rm psr}$ and
  the GW parameters $h_0, \iota, \lambda, \beta, \phi, \omega_{\rm g},
  \phi_{0}$. For a PTA with $N_{\rm psr}$ pulsars, the total number of
  parameters is $N_{\rm psr} + 7$, where the seven represents the GW
  parameters. Details describing the calculation of the ZZ bounds are
  given in Appendix \ref{sec:zzbndapp}. We present the results in
	Figures~\ref{fig:hdep}, where the
  errors of the parameter estimators are given as functions of the GW
  amplitude for various configuration of PTAs. For high-SNR cases, the
  two bounds (ZZ and CR) agree with each other within the
  computational accuracy. For intermediate-SNR cases, the ZZ-bound
  indicates a larger uncertainty than for the CRB predictions. This
  confirms that the CRB underestimates the parameter errors, which is
  mainly due to the multiple solution confusions as we discussed
  before. At the low-SNR limit, the ZZ bound converges to a fixed
  value because of prior information about $\Omega_\eta$, i.e. we made
  use of the fact that the error of the inclination angle $\iota$ and
  source position error will not be larger than $2\pi$, and that the
  error of the GW frequency will not be greater than the sampling
  frequency. From the results, one can also see that the CRB
  underestimates the error of the GW source location and GW frequency
  by one or two orders of magnitudes for intermediate-SNR cases. We
  conclude that the CRB is indeed not an appropriate tool to investigate
  the error performance of PTAs as a GW detector, in accordance with
  the results by \cite{BEv96}.

\begin{figure} \centering \includegraphics[totalheight=3in]{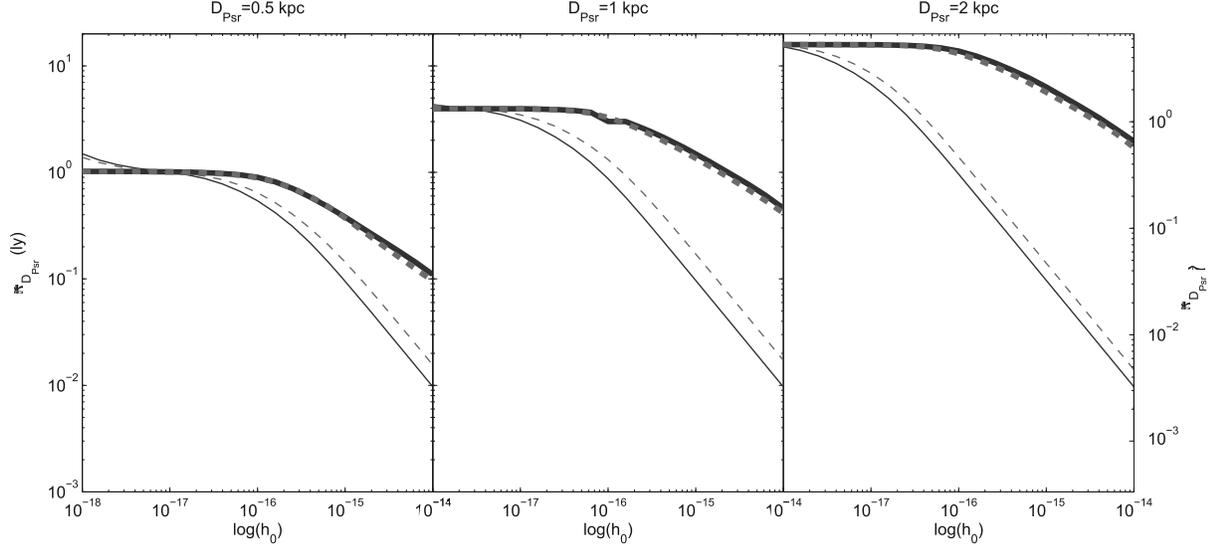}
\caption{Error for pulsar distance measurements as a function of GW
  amplitude. With a PTA GW detection, one can measure the distance to
  the used pulsars with improved accuracy. The
left, middle, and right panel correspond to pulsars with distances of
0.5\,kpc, 1\,kpc, and 2\,kpc, respectively. The thick lines (upper two)
correspond to the ZZ bound and the thin lines (lower two) correspond
to the CRBs. The solid line is for GWs with a period of 2.2\,years and the
dashed lines is for GWs with a period of 4.4\,years. It may be counter
intuitive that the ZZ and CRB agree at the low SNR limit.  However, this is
the consequence of GW-induced multiple solution confusion for pulsar
distances as we discussed in the main text. In fact, the difference
between ZZ bound and the CRB is due to the GW-induced multiple solution
confusion of pulsar distance. Because pulsar timing parallax alone has
well defined unbiased estimators, the ZZ bound and the CRB
  agree with each other when the amplitude of the GW is small and little 
  confusion is introduced. }

\label{fig:dmea} \end{figure}

  We present the error of pulsar distance measurements as a function
  of GW amplitude in Figure~\ref{fig:dmea}.  As a pulsar distance we
  use $0.5-1$\,kpc. The measurements of the pulsar timing
    parallax is for the intermediate-SNR case. Again, the CRB and ZZ
    bounds differ from each other, i.e.\ CRB underestimates the
    measurement errors by orders of magnitude.

\section{Discussion and Conclusions}
\label{sec:con}

In this paper we have shown that we can use a pulsar timing array to
detect the non-evolving single gravitational wave sources and measure
their physical parameters. One can detect a GW source with a GW
amplitude larger than $5\times 10^{-17}$ by timing 20 pulsars
bi-weekly {(once every two weeks)} for 5 years with a timing
accuracy of 15\,ns.  This agrees with the analytical estimation in
equation.~(\ref{eq:estisig}).  Such observational requirements
(i.e.~15\,ns timing accuracy) have not been achieved at present, but
should be possible for future radio-astronomical instruments such as
the Square Kilometer Array (SKA) and possibly, for a limited number of
sources, also for the Five-hundred-meter Aperture Spherical Radio
Telescope (FAST) \citep{NWZZJG04, SLKMSJN09, Kramer10}. {For
  such large telescope sensitivity, radiometer noise will not be the
  limiting factor to achieve a 15 ns-level noise budget.  Rather,
	limiting factors may arise from other physical effects, which we will
  discuss shortly.}

In our treatment of single GW source detection, we do the novel step
of including the pulsar term as an integral part of the signal that
we are looking for. This is possible since the timing parallax offers
the possibility of modelling the pulsar terms accurately for nearby
sources. We take advantage of the pulsar term in several ways. Firstly,
by using the pulsar term, we increase the signal-to-noise ratio (SNR).
{When the pulsar term is completely ignored, the GW signals (only including 
Earth terms) in pulsar timing data are sinusoidal waves with identical phases} 
and the pulsar term becomes a
\emph{noise} source  with the same amplitude compared to the `signal'
term. The SNR of single pulsar data is then always
less than one. {Even worse, the ignored pulsar-term
is sinusoidal signal with unknown phase. It is not a stationary
signal. It is only reduced by averaging over multiple pulsars but not
by time averaging. } On the other hand, if we can use the pulsar term as a
`signal' term, we can sum the signal from multiple pulsar coherently and the SNR 
of single pulsar data can be arbitrarily large,
depending on the GW amplitude.  The second important benefit of utilizing
the pulsar term is that it allows us to get a precise measurement for
the GW source location. The interference between the pulsar term and
the Earth term introduces spiky lobes in the single-pulsar response
pattern (see Figure~\ref{fig:beampat}). In Figure~\ref{fig:skymap},
where the small scale ring-type structures come from individual response
patterns and the large scale quadruple structure is introduced by the
coherence of the Earth term. This large scale quadruple structure is,
in fact, the response pattern, when one only uses the Earth term as a
signal. Clearly the small scale spiky structure increases the locating
ability of PTAs substantially and allows astronomers to search of
electromagnetic counterparts of the GW source.

Thanks to the pulsar timing parallax signal, we are able to include
the pulsar term in our scheme to detect non-evolutionary GW
sources. Our method, first presented in \cite{KS10}, was complemented
recently by the work of \cite{CC10} who also include the pulsar term
in their analysis pipeline. Their approach to use the pulsar term
however differs, as they study the chirped signal and ignore the
timing parallax signal, while we use the pulsar term in combination with
the timing parallax signal.

The pulsar term leads to many potential important applications. For
example, \cite{DF10} recently point out that the curvature of the
wavefronts introduces a GW parallax effect, which can be used to
measure the SMBHB distance. Other possible applications are very
interesting to study also. For instance, while the pulsar distance
measurement helps the GW detection, the GW signal, on the other hand,
also helps the pulsar distance measurement. Due to the interference
between the pulsar term and the Earth term, the GW signal is sensitive
to the pulsar distance. In this way, the detected GW signal can be
used to increase the accuracy of our distance measurements. As we
expected, the overall measurement accuracy for pulsar distances
increases with GW amplitude, as shown in Figure~\ref{fig:dmea}.  The
improvement in accuracy increases faster with respect to the GW
amplitude for pulsars with smaller distances ($D_{\rm psr} = 0.5$\,kpc)
than the ones with larger distances ($D_{\rm psr} = 2$\,kpc). This is
because the distance measurement using the timing parallax has higher
precision for near-by pulsars than for far-away pulsars. A smaller
number of multiple distance solutions occurs for the nearby
pulsars, for which the accuracy of distance measurements increases
faster when the GW amplitude becomes larger.

{In this paper, we have treated the noise as an
additive uncorrelated noise (white noise). Because noise arises
from various different sources: 1) the instrumental noise, e.g.
clock errors, ephemeris errors, polarimetry calibration errors, radio-frequency
interference, etc.; 2) pulsar intrinsic noise, e.g. micro glitch
at even smaller level, switch of spin down rate, pulsar profile change,
etc.; 3) propagation effects,
e.g. DM variation and scattering. Thus the noise is
not necessarily white. For a stochastic GW background detection using pulsar 
timing arrays, a correct treatment of the red noise is a very
important issue, since red noise may destroy any
measurable correlation function \citep{JHLM05}. However the red noise issue is 
less critical for single source detection. This is a consequence of the spectral 
analysis. Since the signal spectrum for non-evolving SMBH will be a single peak 
at the GW frequency, the SNR of the detection is basically the ratio between the 
GW induced signal power and the noise power \emph{in the frequency bin} 
\citep{YHJ10}. Thus, red noise may be dominant at lower frequencies, but less 
effective in the frequency bin of the signal from a single GW source. This is 
fundamentally different from the stochastic GW background detection.

Although several pulsars have timing residuals that are essentially
white \citep{VBC09} at the current level of accuracy, it is still an
unanswered question as to whether the residuals remain white for much
higher timing precision or for much longer observing span. Several authors have 
modeled the
statistics of red noise \citep{FB90, CS10, SC10}, while some works show
the potential of improving timing accuracy by introducing corrections
to interstellar medium effects \citep{HS08, Paul10}. Indeed, for
{ a few} selected pulsars a timing precision at a 30$\sim$50 ns-level 
\citep{Paul10} is
already possible at present, but further studies are needed to see how
and if this can be improved.}

Beside pulsar timing parallax techniques, one can also determine
pulsar distances from other methods, such as the Very Long
Baseline Interferometry (VLBI), pulsar orbital parallax, and relative
acceleration \citep{STWKS10}. Since in this paper we mainly focus on
extracting GW information and pulsar distances from pulsar timing array
data, investigation on incorporating these extra information into PTA GW
data analysis will be presented in future studies.

Finally, we have demonstrated how the errors in the parameter
estimation can be determined reliably.  We have calculated the
parameter estimation error using both the CRB and the ZZ bound. Both
of the methods give nearly identical results in the high SNR region,
as we expected. But the two methods deviate from each other when the
SNR is low. The CRB is known to predict unreachable accuracy in the
low-SNR regime, while the ZZ bound gives more trustful results in
that region.  For most applications involving GW detection, the noise
contributions dominate above the GW signals, so that the ZZ bound
turns out to be superior in reliability compared to the CRB.

\section*{Acknowledgment}

We gratefully acknowledge support from ERC Advanced Grant ``LEAP'', Grant
Agreement Number 227947 (PI Michael Kramer). We thank Fredrick Jenet for his 
suggestions and help. We also thank Alberto Sesana for illuminating discussion 
and Joris Verbiest for reading the manuscript and
his detailed suggestions. We would like to thank Helge
Rottmann for the assistance of using the VLBI cluster to perform most of the
numerical computation in this paper. We also thank the anonymous referee for 
helpful comments.
 
\bibliographystyle{mn2e}
\bibliography{ms}
\clearpage

\appendix

\section{Pulsar timing residuals induced by a monochromatic GW}
\label{sec:2dres}

In this section we consider a non-evolving binary system in a circular orbit as
the source of a monochromatic gravitational wave. If we choose the coordinate
system such that the gravitational wave is in the $z$-axis and the ascending
node along the $x$-axis (reference frame of the source), the two polarizations
can be written as \citep{Blanchet06}
\begin{eqnarray}
  h_+(t)      &=& h_0 \cos(\omega_{\rm g} t + \Phi_0)\,\frac{1 +\cos^2\iota}{2} \;, 
              \label{eq:hp} \\ 
  h_\times(t) &=& h_0 \sin(\omega_{\rm g} t + \Phi_0)\,\cos\iota \;,
              \label{eq:hc}
\end{eqnarray}
$\Phi_0$ is a constant phase shift and $\iota$ denotes the angle between the
orbital angular momentum of the source and the direction from the source to the
Earth. The amplitude $h_0$ depends on the binary parameters and distance (see
Section~\ref{sec:timres}). For a pulsar at distance $D_{\rm psr}$,
with azimuthal angle $\psi$ and with polar angle $\theta$,  the residuals caused
by the gravitational wave are given by \citep{EW75, Detweiler79, Wahlquist87}
\begin{equation}
  R(t) = (1 + \cos\theta) \,\left[
		   (H_+^E      - H_+^p)\cos(2\psi) + 
		   (H_\times^E - H_\times^p)\sin(2\psi) \right] \;.  
\end{equation}
where
\begin{eqnarray}
  H_+^E - H_+^p &=& 
    \frac{h_0}{2\omega_{\rm g}}\sin(\Delta\Phi/2)(1 + \cos^2\iota)
    \nonumber\\ && \qquad
	\left[ \cos(\omega_{\rm g} t + \Phi_0 - \Delta\Phi/2) -
	       \cos(\Phi_0 - \Delta\Phi/2) \right] \;,\\ 
  H_\times^E - H_\times^p 
			&=& \frac{h_0}{2\omega_{\rm g}} \sin(\Delta\Phi/2)(2\cos\iota)
    \nonumber\\ && \qquad
	\left[ \sin(\omega_{\rm g} t + \Phi_0 - \Delta\Phi/2) -
		\sin(\Phi_0 - \Delta\Phi/2) \right] \;,
\end{eqnarray}
where
\begin{equation}
  \Delta\Phi = \omega_{\rm g} D_{\rm psr} (1 - \cos\theta) \;. 
\end{equation}
After dropping constant terms, the timing residuals caused by the
gravitational wave reads
\begin{eqnarray}
  R(t) &=& \frac{h_0}{2\omega_{\rm g}}(1 + \cos\theta)\sin(\Delta\Phi/2) 
  \nonumber\\
       &&  \left[
		   (1 + \cos^2 \iota) \cos(2\psi) \cos(\omega_{\rm gw} t + \Phi_0') +             
		   2 \cos\iota        \sin(2\psi) \sin(\omega_{\rm gw} t + \Phi_0') 
		   \right] \;,
		   \label{eq:2dtim}
\end{eqnarray}
where $\Phi'= \Phi_0 - \Delta\Phi/2$. One can easily convert these results to 
the representation in ecliptic coordinates as used in the main text. The 
representation here has the merit of clearly presenting the physical pictures.  
For example, it looks like a singularity at $\theta = 0$ in 
equation.~(\ref{eq:redshiftsig}), but the representation here, 
equation.~(\ref{eq:2dtim}), clearly  shows that the singularity is canceled by 
geometrical factors of GW polarization. The amplitude modulation ($\sin(\Delta 
\Phi/2)$) and phase modulation ($\Delta \Phi/2$) can be easily understood
using the identity $e^{{\rm i}(\omega t + \Delta\Phi)} - e^{ {\rm i}\omega t}
= 2 \sin(\Delta\Phi/2) e^{{\rm i}(\omega t + \Delta\Phi/2+\pi/2)}$, where
the interference between two sinusoidal signals with a phase difference of
$\Delta\Phi$ introduces an amplitude modulation of $\sin(\Delta\Phi/2)$ and
a phase shift of $\Delta\Phi/2 + \pi/2$.

One can also calculate the results using the observer frame.  The waveform of 
GWs from a SMBHB with a circular orbit at
cosmological redshift $z$ and chirp mass $\cal M$ is given by \citep{Wahlquist87, 
Blanchet06, Hughes09}
\begin{equation}
  h_{ij} = \epsilon_{ij}^{+}A_{+} + \epsilon_{ij}^{\times}A_{\times} \;,
  \label{eq:gwform}
\end{equation}
where the $A_{+}$ and $A_{\times}$ are the waveforms of the two polarization 
modes `+' and `$\times$', respectively. The polarization tensors 
$\epsilon_{ij}^{+}$ and $\epsilon_{ij}^{\times}$ are \citep{Wahlquist87}
\begin{eqnarray}
\epsilon_{ij}^{+} &=&
\left(
\begin{array}{ccc}
   \sin^2\lambda - \cos^2\lambda  \sin^2\beta            & 
  -\sin\lambda \cos\lambda \left(\sin ^2\beta + 1\right) & 
   \cos\lambda \sin\beta \cos\beta                       \\ 
  -\sin\lambda \cos\lambda \left(\sin ^2\beta + 1\right) & 
   \cos^2\lambda - \sin^2\lambda  \sin^2\beta            & 
   \sin\lambda \sin\beta  \cos\beta                      \\
   \cos\lambda \sin\beta  \cos\beta                      & 
   \sin\lambda \sin\beta  \cos\beta                      & 
  -\cos ^2\beta 
\end{array}
\right) \;, 
\\
\epsilon_{ij}^{\times} &=&
\left(
\begin{array}{ccc}
   \sin(2\lambda) \sin\beta & 
  -\cos(2\lambda) \sin\beta & 
  -\sin\lambda    \cos\beta \\
  -\cos(2\lambda) \sin\beta & 
  -\sin(2\lambda) \sin\beta & 
   \cos\lambda    \cos\beta \\
  -\sin\lambda    \cos\beta & 
   \cos\lambda    \cos\beta & 
   0
\end{array}
\right) \;,
\label{eq:polarten}
\end{eqnarray}
where $\lambda$ and $\beta$ are the ecliptic longitude and latitude of the 
GW source. For a non-evolving SMBHB with circular orbit, the GW waveform 
$A_{\times}$ and $A_{+}$ of the two polarization modes are \citep{Wahlquist87}
\begin{eqnarray}
  A_{+} &=&
     h_0 \left[ \cos\iota \sin(2\phi) \sin(\omega_{\rm g}t )  
              - \frac{1}{2}(1 + \cos^2\iota) 
                \cos(2\phi) \cos(\omega_{\rm g} t) \right] \;, 
  \label{eq:ampgw1}\\ 
  A_{\times} &=&
    -h_0 \left[ \cos\iota  \cos(2 \phi) \sin(\omega_{\rm g} t)  
							 + \frac{1}{2}(1 + \cos^2\iota) \sin(2\phi) \cos(\omega_{\rm g} 
							 t) \right] \;.
	\label{eq:ampgw}
\end{eqnarray}
The angle $\iota$ is the orbital inclination of the 
GW binary source, i.e.\ the angle between orbital angular momentum and the 
gravitational wave vector. The angle $\phi$ defines the direction of
the binary ascending node on the sky. The GW source chirp mass $\cal M$ and the 
source distance $D_{c}$ degenerate to the GW amplitude $h_0$, if one only has 
the PTA data.  Instead of discussing the detection statistics using full GW 
source parameters (${\cal M}, D_{c}, \omega_{\rm g}$), it is much better to use 
the reduced parameter sets ($h_0, \omega_{\rm g}$), which will be used from now
on.

A GW affects the pulsar timing by introducing shifts to the observed rotational 
frequency of the pulsar, $\nu$ (see refs.~\cite{Detweiler79, LJR08}, and 
Appendix~\ref{sec:2dres} for details)
\begin{equation}
	\frac{\delta\nu}{\nu} = -\frac{1}{2} \, \frac{\hat{n}_{i} \hat{n}_{j}}{1 - 
	\cos\theta}
     \left[h_{ij}(t,0) - h_{ij}(t - D_{\rm psr}, {\bf D}_{\rm psr})\right] \;,
  \label{eq:redshiftsig}
\end{equation}
where $\hp$ is the unit vector pointing from the observer to the pulsar 
and $\theta$ is the angle between the GW source direction and the pulsar 
direction for the observer. The
$h_{ij}(t,0)$ is the GW strain at the Earth (the introduced \emph{Earth term}) and the
$h_{ij}(t - D_{\rm psr}, {\rm \bf D_{\rm psr}})$ is the GW strain at the pulsar 
(the introduced \emph{pulsar term}). The explicit forms for  $\hp$ and the $\cmu$ are
\begin{eqnarray}
  {\rm \bf \hp}&=&
  \left(
  \begin{array}{c}
    \cos\lambda_{\rm p} \cos \beta_{\rm p} \\
    \sin\lambda_{\rm p} \cos \beta_{\rm p} \\
    \sin\beta_{\rm p}
  \end{array}
  \right) \;, 
  \\
  \cmu &=&
    \cos\beta \cos\beta_{\rm p} \cos(\lambda - \lambda_{\rm p}) + 
    \sin\beta \sin\beta_{\rm p}
  \label{eq:psrdef}
\end{eqnarray}
where $\lambda_{\rm p}, \beta_{\rm p}$ are again the ecliptic longitude and latitude for
the pulsar position.

The GW induced TOA signal $R_{\rm g}(t)$ is the integration of the frequency
shift over the observer's local time, $R_{\rm g}(t) = \int
\frac{\delta\nu(t)}{\nu} \,dt$ \citep{JHLM05}. Inserting 
equation.~(\ref{eq:ampgw1})
and equation.~(\ref{eq:ampgw}) into Equation.~(\ref{eq:gwform}) and integrating
equation.~(\ref{eq:redshiftsig}) over time yields Equation,~(\ref{eq:gwsig}).

\section{Analytical calculation for the errors of timing parallax
measurements} \label{sec:estipar}

In this section, we derive an analytic expression for the uncertainty
in the distance measurement for pulsars using the timing parallax. As a
first-order approximation, we assume that the Earth follows a circular
orbit. In practice, the Earth orbit is not only eccentric, but also
shows additional, quite complex deviations from an elliptical orbit
due to perturbations by other masses in the Solar System. However,
because the eccentricity of the Earth orbit is small, and we only need
a first order estimation of the error in the parallax measurement, our
assumption is justified. A more complete discussion on pulsar timing
parallax measurements can be found in the literature \citep{BH86, RT91, LK05}.

We assume that the pulsar TOA signal is composed of the timing parallax term
and a white Gaussian noise term representing noise contributions from 
other processes, i.e.  
\begin{equation}
  R(t) = R_{\rm par}(t)+n(t) \;.  
  \label{eq:resusig} 
\end{equation}
Regarding the time series of signals as a vectors, e.g.\ ${\rm \bf R}=R(t_i)$, 
$i=1\ldots N_{\rm obs}$, where the $N_{\rm obs}$ is the number of data points
and $t_i$ is the time of the $i$-th observation. The probability
distribution, $f({\rm \bf R}| D_{\rm psr})$, for the TOA data $\rm \bf R$,
with given pulsar distance $D_{\rm psr}$, is
\begin{equation}
  f({\rm \bf R}|D_{\rm psr}) = 
	\left(\frac{1}{\sqrt{2\pi}\sigma_{\rm n}}\right)^{N_{\rm obs}} {\rm 
	Exp}\left[-\sum_{i=1}^{N_{\rm obs}} \frac{\left[R(t_i) - R_{\rm 
	par}(t_{i})\right]^2}
		 {2\sigma_{\rm n}^2}\right]\;.
  \label{eq:probr}
\end{equation}
Using the CRB, the uncertainty of pulsar distance measurements is
\begin{equation}
  \sigma_{D_{\rm psr}}^2\le\frac{1}{{\int}\!  \left[\frac{ \mathrm{d}
	\ln f({\rm \bf R}|D_{\rm psr})}{\mathrm{d} D_{\rm psr}}\right]^2 f({\rm
    \bf R}|D_{\rm psr}) \, \mathrm{d}{\rm \bf R} } \;, 
  \label{eq:CRbound}
\end{equation}
which corresponds to equation.~(\ref{eq:distmeasure}) of the main text.  We have 
checked this result against simulations with the pulsar timing
software TEMPO, which uses the planetary ephemeris DE405 for a highly
accurate representation of the Earth orbit in the Solar System barycenter.  For 
pulsars which are not too close to the poles of the ecliptic (more than
one degree angular distance), our analytic equation (\ref{eq:probr}) agrees
to better than a few percent. 

\section{Ziv-Zakai bounds for array signal with additive noise}
\label{sec:zzbndapp}

The ZZ bound was first introduced for a scalar parameter estimation
problem by \cite{CZZ75}, and was later extended to general vector
parameter estimation problems \citep{BSE94}. Its first use for GW
studies was by \cite{NV97}. In this paper, we only summarize the key
results of the ZZ bound and refer readers to the literatures cited for
further information.

We suppose that we are dealing with timing data from an array of $N_{\rm psr}$ 
pulsars. The data of each pulsar is a $N$-point TOA serial signal. We denote the 
whole timing data set by $R_j^i$, where the superscript is the pulsar index, and 
the subscript is the data index. For example, the $R_j^i$ is the $j$-th data 
point of the $i$-th pulsar TOA. The timing data is just a $N_{\rm 
psr}$-times-$N$ dimensional vector ${\rm \bf R}$ and their components 
$R_{j}^{i}$ takes the form of equation.~(\ref{eq:fullsig}).  The
signal part of $\rm \bf R$ is completely determined by the $k$-dimensional 
parameter $\lambda_k$ (GW parameters plus pulsar parameters). The noise part is
a random variable following independent Gaussian statistics. We have
\begin{equation}
  R_j^i=s_j^i(\lambda_k)+n_j^i,
  \label{eq:sigmod}
\end{equation}
where the signal part ${\rm \bf s}={\rm \bf R}_{\rm g}+{\rm \bf R}_{\rm par}$, 
and the noise $\rm \bf n$ is assumed to be an un-correlated white Gaussian 
noise, i.e. $\langle n_j^i n_j^i\rangle=\sigma_{\rm n}^{2\;(i)}$ and $\langle 
n_{j'}^{i} n_j^{i'}\rangle=0$ for any $j'\neq j$ or $i' \neq i$.  The  
$\sigma_{\rm n}^{(i)}$ is the RMS level for the $i$-th pulsar TOA noise from 
other contributions.  The $P_{\epsilon, {\rm min}}(\lambda_{k}, \lambda_{k}')$ 
is the minimal probability of making a mistake in deciding which parameter set, 
the $\lambda_{k}$
or $\lambda_{k}'$, to choose, when giving data ${\rm \bf R}$. For the 
case where $\rm \bf \lambda$ and $\rm \bf \lambda'$ have equal prior, as shown 
by \cite{Kassam88}, the following likelihood ratio test minimizes the probability of 
making mistakes,
\begin{equation}
  \textrm{choose }\left\{ \begin{array}{c}
	\lambda_{k} \textrm{, if } \frac{f({\rm \bf R}|\lambda_{k})}{f({\rm \bf 
	R}|\lambda_{k}')}>1 \\
	\lambda_{k}' \textrm{, if } \frac{f({\rm \bf R}|\lambda_{k})}{f({\rm \bf 
	R}|\lambda_{k}')}<1
  \end{array}
  \right.,
  \label{eq:minalg}
\end{equation}
where the $f({\rm \bf R}|\lambda_{k})$ is the probability distribution function
of the data ${\rm \bf R}$, when the parameters take value of $\lambda_{k}$.  
Following the likelihood test in equation.~(\ref{eq:minalg}), one can show
\begin{equation}
  P_{\epsilon, {\rm min}}(\lambda_{k}, 
  \lambda_{k}')=\frac{1}{2}\left[P\left(\left.\frac{f({\rm \bf 
  R}|\lambda_{k})}{f({\rm \bf 
  R}|\lambda_{k}')}>1\right|\lambda_k'\right)+P\left(\left.\frac{f({\rm \bf 
  R}|\lambda_{k})}{f({\rm \bf R}|\lambda_{k}')}<1\right|\lambda_k 
  \right)\right].
  \label{eq:errprob}
\end{equation}
For the uncorrelated white Gaussian noise,
the likelihood ratio $f({\rm \bf R}|\lambda_k)/f({\rm \bf R}|\lambda_k')$ is
\begin{equation}
	\left. \frac{f({\rm \bf R}| \lambda_k)}{f({\rm \bf 
	R}|\lambda_k')}\right|_{\lambda_{0,k}}=\frac{ {\rm 
	Exp}\left[{-\sum_{i=1}^{N_{\rm psr}}\sum_{j=1}^{N}
	\left(s_j^i(\lambda_{0,k})-s_j^i(\lambda_{k})+n_j^i\right)^2/\left(2\sigma_{n}^{2\;(i)}\right)}\right]}{ 
	{\rm Exp}\left[{- \sum_{i=1}^{N_{\rm 
	psr}}\sum_{j=1}^{N}\left(s_j^i(\lambda_{0,k})-s_j^i(\lambda_{k}')+n_j^i\right)^2/\left(2\sigma_{n}^{2\; 
	(i)}\right)}\right]}.
  \label{eq:likrat}
\end{equation}
Integrating over the random vectors $n_{j}^i$, one derives
\begin{equation}
  P\left(\left.\frac{f({\rm \bf R}|\lambda_{k})}{f({\rm \bf 
  R}|\lambda_{k}')}>1\right|\lambda_{0,k}\right)=\frac{1}{2}{\rm Erfc}\left[- 
  \frac{ \sum_{i=1}^{N_{\rm psr}}\frac{1}{\sigma_{\rm n}^{2 \; i}}\sum_{j=1}^N 
  (s_j^i(\lambda_{0,k})-s_{1}^i)(s_{j}^i(\lambda_k)-s_{j}^i(\lambda_k'))}{\sqrt{2} 
  \sqrt{ \sum_{i=1}^{N_{\rm psr}} \frac{1}{\sigma_{\rm n}^{2\; 
  (i)}}\sum_{j=1}^N\left(s_j^i(\lambda_k)-s_{j}^i(\lambda_k')\right)^2}}\right],  
  \label{eq:errprob1}
\end{equation}
and the $P\left(\left.\frac{f({\rm \bf R}|\lambda_{k})}{f({\rm \bf 
R}|\lambda_{k}')}<1\right|\lambda_{0,k} \right)$ can be evaluated in very 
similar fashion. The minimal error probability  $P_{\epsilon, {\rm min}}$ is 
then
\begin{equation}
  P_{\epsilon, {\rm min}}(\lambda_{k}, \lambda_{k}') =
  \frac{1}{2}{\rm Erf}(\zeta),
  \label{eq:errprobfinal}
\end{equation}
where
\begin{equation}
  \zeta =
  \sqrt{ \sum_{i=1}^{N_{\rm 
  psr}}\frac{\sum_{j=1}^N\left[s_{j}^i(\lambda_k)-s_{j}^i(\lambda_k')\right]^2}
  {2\,\sigma_{\rm n}^{2 \; (i)}}}.
\label{eq:beta}
\end{equation}
equation.~(\ref{eq:beta}) and (\ref{eq:errprobfinal}) show that the larger 
difference between the signal of $\lambda_{k}$ and the signal of $\lambda_{k}'$, 
the larger $\zeta$ is, and the less probability one will make
a mistake in discriminating between the two signals.
\section{Generalization of the amplitude condition to the red noise case} 
\label{sec:genred}
{
In the main text, equation~(\ref{eq:estisig}) is valid only for white-nosie.  
Since equation~(\ref{eq:estisig}) is quite useful to estimate the detectable 
amplitude of monochromatic GWs, we generalize it here also for the red-noise 
case.  The easiest way of deriving equation~(\ref{eq:estisig}) is by using a 
spectral analysis in frequency space.

We focus on the frequency bin, which contains the GW signal, i.e. the frequency 
band of $[f-\frac{1}{2} \Delta f, f+\frac{1}{2} \Delta f]$. For signals from 
$N_{\rm psr}$ pulsars, the total GW signal power $P_{\rm s}$ in the frequency 
bin is
\begin{equation}
	P_{\rm s}=\left(\frac{ N_{\rm psr}\overline{h}}{2\omega_{\rm g}}\right)^2\,,
\end{equation}
and the noise power is \begin{equation}
	P_{\rm n}= N_{\rm psr} S_{n}(f) \Delta f= \frac{ N_{\rm psr} S_{n}(f) f_{\rm 
	max}}{N_{\rm obs}}\,,
\end{equation}
where the $S_{n}(f)$ is the spectra of $n(t)$, the noise component. The $f_{\rm 
max}$, defined in Section~\ref{sec:esti}, is the bandwidth of the signal.  The 
detectable condition is then
$P_{\rm s}\ge P_{\rm n}$, which leads to
\begin{equation}
	\frac{\overline{h}}{2\omega_{\rm g}}\ge \sqrt{\frac{S_{\rm n}(f) f_{\rm 
	max}}{N_{\rm psr}N_{\rm obs}}}\,.
	\label{eq:geestisig}
\end{equation}
In this way, the generalization of equation~(\ref{eq:estisig}) to the red-noise 
case is simply to replace the RMS noise level $\sigma_{\rm n}$ with the value 
of $\sigma_{\rm e}=\sqrt{S_{\rm n}(f) f_{\rm max}}$, the effective level of 
noise. 	}
\label{lastpage}

\end{document}